\documentclass[reprint,amsmath,amssymb,aps,longbibliography]{revtex4-2}

\usepackage{graphicx,amsmath,amssymb,color}

\usepackage{comment}

\begin{document}

\title{Anomalous weak values via a single photon detection}

\author{E. Rebufello$^{1,2}$, F. Piacentini$^1$, A. Avella$^1$, M. A. de Souza$^{3}$, M. Gramegna$^1$, J. Dziewior$^{4,5}$, E. Cohen$^6$, L. Vaidman$^7$, I. P. Degiovanni$^1$, M. Genovese$^1$}
\affiliation{$^{1}$INRIM, Strada delle Cacce 91, I-10135 Torino, Italy}
\affiliation{$^{2}$Politecnico di Torino, corso Duca degli Abruzzi 24, I-10129 Torino, Italy}
\affiliation{$^{3}$National Institute of Metrology, Quality and Technology - INMETRO; Av. Nossa Senhora das Gra\c{c}as, 50, 25250-020, Duque de Caxias- RJ– Brazil}
\affiliation{$^4$Max-Planck-Institut f\"{u}r Quantenoptik, Hans-Kopfermann-Stra{\ss}e 1, 85748 Garching, Germany}
\affiliation{$^5$Department f\"{u}r Physik, Ludwig-Maximilians-Universit\"{a}t, 80797 M\"{u}nchen, Germany}
\affiliation{$^6$Faculty of Engineering and the Institute of Nanotechnology and Advanced Materials, Bar Ilan University, Ramat Gan 5290002, Israel}
\affiliation{$^7$Raymond and Beverly Sackler School of Physics and Astronomy, Tel-Aviv University, Tel-Aviv 6997801, Israel}
\affiliation{$^\ast$To whom correspondence should be addressed; E-mail: f.piacentini@inrim.it}

\date{}

\maketitle

\begin{figure}
  \centering
  \includegraphics[width=0.48\textwidth]{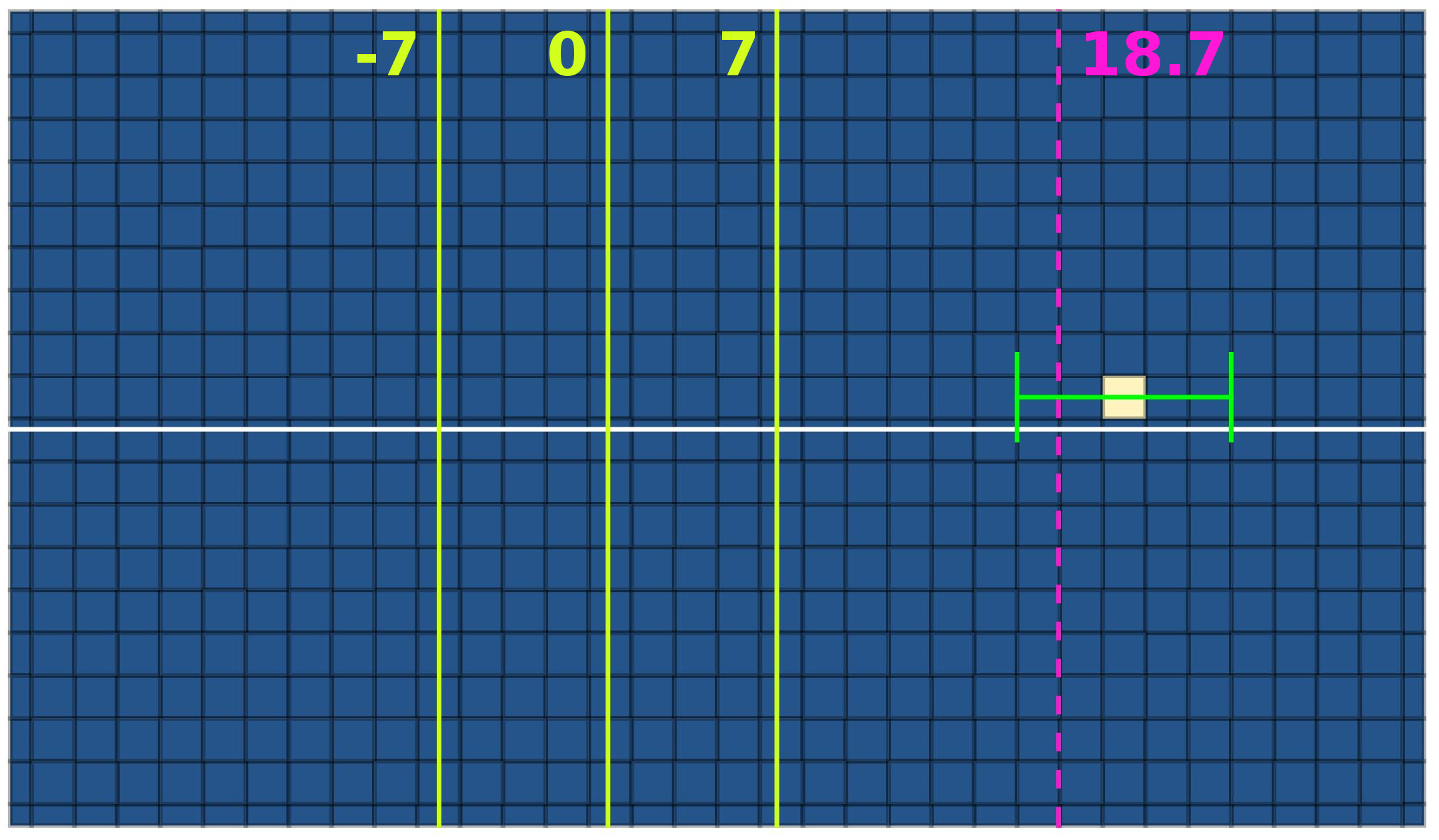}
	\caption{\textbf{Single detection event yielding an anomalous weak value of $ \sigma^\Sigma_3$.}
	The vertical solid lines show the borders and center of the eigenvalue spectrum, while the dashed line indicates the weak value calculated according to the experimental parameters, i.e. $( \sigma^\Sigma_3)_w=18.7$.
	The experimental point, shown in white, gives the value $(\sigma^\Sigma_3)^{\text{1~click}}_w=21.4$. The uncertainty, represented by the horizontal green bar, is specified by calculating the width of the photon wave function before the detection and confirmed by repeating the experiment many times (see Fig.~\ref{WV_contour_single}).
    }\label{anomalous}
\end{figure}

\textbf{Is it possible that a measurement of a spin component of a spin-1/2 particle yields the value 100?
In 1988 Aharonov, Albert and Vaidman argued that upon pre- and postselection of particular spin states, weakening the coupling of a standard measurement procedure ensures this paradoxical result \cite{AAV}.
This theoretical prediction, called \textit{weak value}, was realized in numerous experiments
~\cite{Ex1,Ex2,Ex4,WhiteLight,pusey2,LundNat,GoggPNAS,A4}, but its meaning remains very controversial 
\cite{ContrLeggett,ContrPeres,ReplyAV,ContrFerrie,JordanWA,PangEntWA,ContrPusey,Pro1,PiacentiniInt}, since its ``anomalous'' nature, i.e. the possibility to exceed the eigenvalues range, as well as its ``quantumness'' are debated \cite{FerrieCoin,MundarainWV,vaidmanBeyond}. 
We address these questions by presenting the first experiment measuring anomalous weak values with just a single click, without any statistics. The measurement uncertainty is significantly smaller than the gap between the measured weak value and the nearest eigenvalue. Beyond clarifying the meaning of weak values, this result represents a breakthrough in understanding quantum measurement foundations, paving the way to further applications of weak values to quantum photonics.}
Weak values (WVs) \cite{AAV} represent one of the most interesting and intriguing new quantum measurement paradigms.
In the paper introducing them \cite{AAV}, the maximal eigenvalue (in appropriate units) was 1, but the WV of the measured spin component was 100.
Weakening the coupling in the measurement procedure made the uncertainty in an individual measurement much larger than 1 (and even than 100), thus this ``anomalous'' value was observed only after averaging over a very large number of readings of the pointer variable.
While averaging is a standard practice in many measurement protocols, postselection is not, hence the legitimacy of the statistical analysis was questioned \cite{FerrieCoin,MundarainWV,vaidmanBeyond}. 
Understanding this matter is fundamental not only for clarifying the meaning of WVs, but also in view of significant applications in quantum metrology \cite{Ex2,Ex4,WhiteLight,Ex5,LyoPRL}. 
In this work we present a {\it robust weak measurement} -- an experiment in which a single reading of the measuring device, coupled to the system only once, provides a WV and, in particular, an anomalous one.
Postselection still plays a crucial role, but the anomalous outcome no longer arises from a statistical analysis.
We measured an observable with eigenvalues in the range $[-7,7]$. The WV of the observable of the pre- and postselected system on which a single-click measurement was performed was 18.7, and our single click yielded $21.4\pm4.5$, see Fig.~\ref{anomalous}.
This is a surprising result, since the expectation value of the observable in the preselected state was only $2.2$.
It would not be surprising that postselection on the eigenstate corresponding to a maximal eigenvalue (i.e. 7) slightly increased the measured value, but only up to 7, not beyond (in fact, in our experiment the expectation value corresponding to the postselected state was also just $2.2$).\\
\begin{figure}
  \centering
  \includegraphics[width=0.45\textwidth]{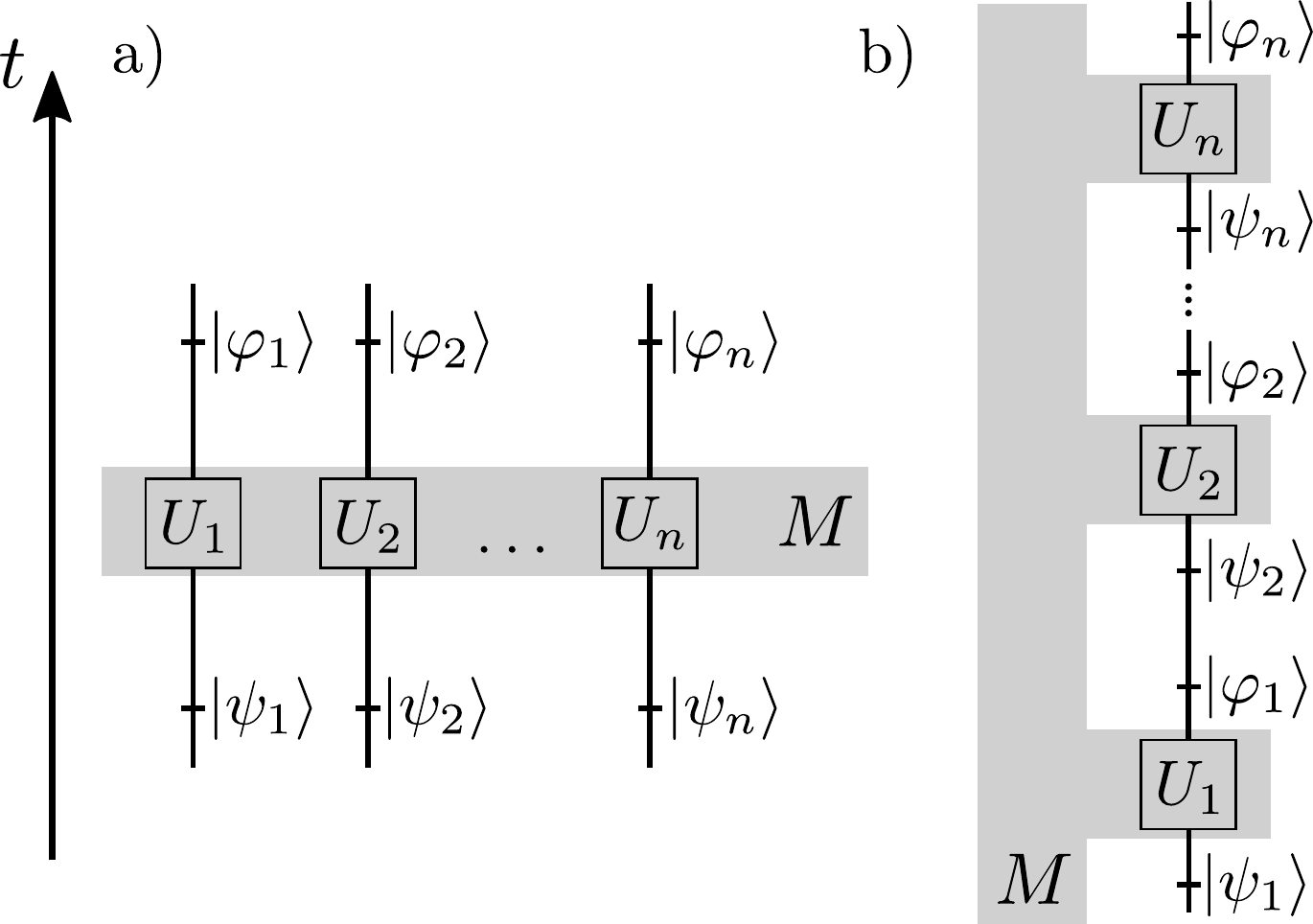}
    \caption{\textbf{Robust weak measurement: theoretical framework.}  a) A measuring device $M$ is coupled simultaneously to $n$ particles of a pre- and postselected  system.  b) The measuring device is coupled to the same particle at $n$ times with particular pre- and postselection at each time.} \label{figAnalogy}
\end{figure}
The main theoretical basis for our experiment is the work \cite{Surp}, which preceded the introduction of WVs \cite{AAV}.
Today, after the development of the WV formalism, the easiest way to explain this work is via weak measurements performed on $n$ particles with a single measuring device, see Sec. VII of \cite{AV90}.
Within the standard weak measurement procedure, in which each particle has its own measuring device, the uncertainty of measuring the sum of variables related to all particles scales like $\sqrt{n}$.
In our modified procedure for measuring $A^\Sigma\equiv \sum_{k=1}^n A_k$, being $A_k$ an observable of the $k$-th system, we couple a single measuring device to all particles and thus, importantly, the uncertainty doesn't increase with the number of particles.
This allows extracting an anomalous WV $(A^\Sigma)_w$ by coupling only once to a single system and observing only a single click of the measuring device.
In the present experiment, instead of a system consisting of several particles, all preselected at the beginning and postselected after the weak couplings, Fig.~\ref{figAnalogy}a, we consider a single particle coupled to a measuring pointer at several times in-between a series of separate pre- and postselections, Fig.~\ref{figAnalogy}b.
These two scenarios result in the same effect on the measuring device, but our scheme is much easier to implement.
The observable, i.e. the sum of polarization variables of $n$ photons, is replaced by the sum of polarization variables of the same photon at $n$ different times, while the spatial degree of freedom of this single photon plays the role of the measuring device.
The measured variable is
\begin{align}
 \sigma^\Sigma_3 \equiv \sum_{k=1}^n \sigma^{(k)}_3,
\end{align}
where $\sigma_3\equiv|H\rangle\langle H|-|V\rangle\langle V|$.
To measure the WV of $\sigma^\Sigma_3$ we introduce a sequence of $n=7$ weak couplings sandwiched between pre- and postselections determined by linear polarization filters and polarization rotators (see Methods).
The preselected state $|\psi_\alpha\rangle$ and the postselected state $|\psi_\beta\rangle$ are
\begin{subequations}
\begin{align}
|\psi_\alpha\rangle &= \cos\alpha |H\rangle + \sin\alpha |V\rangle, \\
|\psi_\beta\rangle &= \cos\beta |H\rangle + \sin\beta |V\rangle.
\end{align}
\end{subequations}
At each stage, we induce an effective interaction Hamiltonian
\begin{align}
 H=g(t)\,\sigma_3 \otimes p_x,
 \end{align}
where $p_x$ is the photon transverse momentum and $g(t)$ characterizes the coupling strength.
The WV is measured via an accumulated shift along the $x$ axis of the spatial wave function of the photon, initially described by a Gaussian:
\begin{equation}\label{chi0}
\chi(x) = \frac{1}{\sqrt{\Delta\sqrt{2\pi}}}e^{-x^2/4\Delta^2}.
\end{equation}
Note that we calibrate the physical measurement device to provide a direct WV reading such that $x$, $g(t)$ and $\Delta$ are dimensionless and properly re-scaled with $\int g(t)\,\mathrm{d}t=1$.
We have to make the coupling strong enough such that the deviation of the WV from the closest non-anomalous value becomes larger than the uncertainty on the WV itself.
This strong coupling affects the state evolution and makes the WV less anomalous \cite{PiacentiniInt,Univ}, hence we need to choose our parameters carefully for a conclusive demonstration of the effect.
Even in presence of a non-negligible coupling to the measuring device, the system has well-defined WVs for every observable at every moment between pre- and postselection.
In general, however, the measuring device doesn't indicate this WV, because of the entanglement between system and measuring device, see Fig.~5 from \cite{vaidmanBeyond}.
Somewhat surprisingly, in the case of coupling to a Gaussian pointer, the expectation value of the final pointer exactly equals this WV.
The polarization WV is constant in time during the measurement interaction, so we can calculate it at the moment just before the postselection, when the composite polarization and spatial state of the photon is
\begin{align}
\cos\alpha|H\rangle|\chi_{+}\rangle+\sin\alpha|V\rangle|\chi_{-}\rangle,
\end{align}
where $|\chi_{+}\rangle$ and $|\chi_{-}\rangle$ denote Gaussians shifted by $1$ and $-1$ respectively.
Thus, the photon has a mixed polarization state described by the density matrix $\rho_\alpha$, that can be expressed in the $\left\lbrace |H\rangle,|V\rangle \right\rbrace$ basis as
\begin{align}
\rho_\alpha &= \begin{pmatrix}
\cos^2 \alpha & e^{-\frac{1}{2\Delta^2}} \sin \alpha \cos \alpha \\
e^{-\frac{1}{2\Delta^2}} \sin \alpha \cos \alpha & \sin^2 \alpha
\end{pmatrix}. \end{align}
Postselecting on $| \psi_\beta \rangle$, the WV is given by (see Eq.~(32) from \cite{vaidmanBeyond})
\begin{align}\label{WV1_eta}
(\sigma_3)_w &= \frac{\mathrm{tr} \left( | \psi_\beta \rangle \langle \psi_\beta | \sigma_3 \rho_\alpha \right)}{\mathrm{tr} \left( | \psi_\beta \rangle \langle \psi_\beta | \rho_\alpha \right)}
=  \frac{\mu^2-\nu^2}{\mu^2+ \nu^2+2\mu\nu e^{-\frac{1}{2\Delta^2}}},
\end{align}
where $\mu = \cos\alpha\cos\beta$ and $\nu = \sin\alpha\sin\beta$.

To evaluate the expected result in the case of $n$ couplings, we use the equivalence between our system, Fig.~\ref{figAnalogy}b, and the scenario of $n$ pre- and postselected systems coupled to the same pointer, Fig.~\ref{figAnalogy}a.
We define the joint states of the $n$ systems as $|\Psi_\alpha \rangle = \bigotimes^n_{k=1} |\psi^{(k)}_\alpha\rangle$ and ${|\Psi_\beta \rangle = \bigotimes^n_{k=1} |\psi^{(k)}_\beta\rangle}$.
The WV is
\begin{align}\label{WV7_eta}
\left(\sigma^\Sigma_3\right)_w &= \frac{ \langle\Psi_\beta| \sigma^\Sigma_3 \mathrm{tr}\left( U^{\Sigma} |\Psi_\alpha\rangle \langle\Psi_\alpha| \otimes |\chi\rangle \langle \chi | (U^\Sigma)^\dagger\right) |\Psi_\beta\rangle }{\langle\Psi_\beta| \mathrm{tr}\left( U^{\Sigma} |\Psi_\alpha\rangle \langle\Psi_\alpha| \otimes |\chi\rangle \langle \chi | (U^\Sigma)^\dagger\right) |\Psi_\beta\rangle} \nonumber \\
&= \frac{\sum_{k,l=0}^{n}\binom{n}{k}\binom{n}{l}\mu^{k+l}\nu^{2n-k-l} \left(2k-n\right)\gamma_{kl}} {\sum_{k,l=0}^{n}\binom{n}{k}\binom{n}{l}\mu^{k+l}\nu^{2n-k-l}\gamma_{kl}},
\end{align}
where $U^{\Sigma}=e^{-i \sum^n_{k=1} \sigma^{(k)}_3 \otimes p_x}$, $\gamma_{kl}=e^{-\frac { (k-l)^2}{2\Delta^2}}$, and the trace is taken over the pointer system only.
%
\begin{figure}
  \centering
  \includegraphics[width=0.45\textwidth]{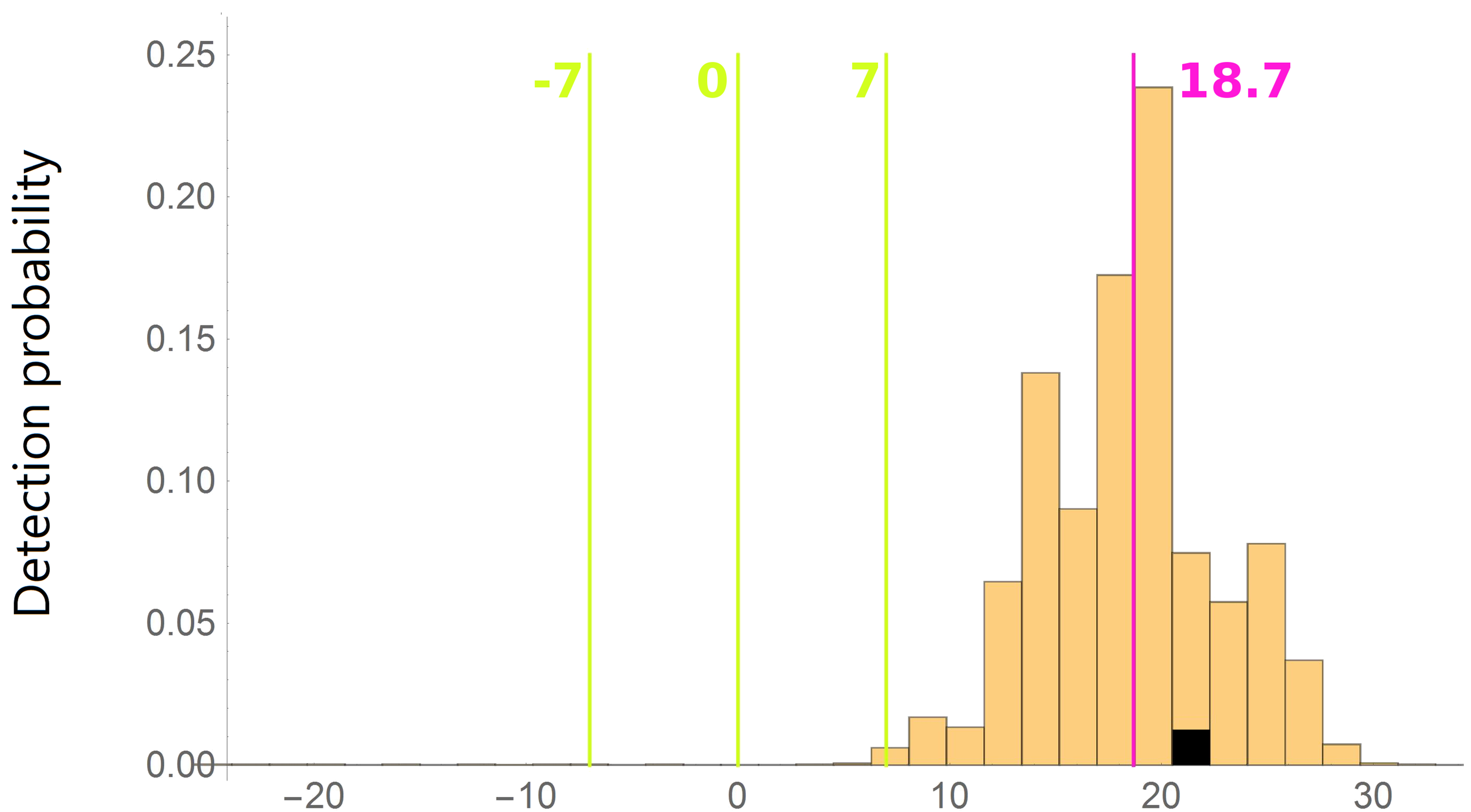}
  \caption{\textbf{Measurement of anomalous weak value.} Normalized histogram of the photon counts along the $x$ axis of the EM-CCD for repetitions of the single-click experiment (with unchanged parameters).
  The black square indicates the first click of the run, corresponding to the single-click experiment. The light green lines indicate the borders and center of the eigenvalue spectrum. The purple line shows the expected (theoretical) weak value $(\sigma^\Sigma_3)_w$.
  }
  \label{WV_contour_single}
\end{figure}
The single photon click presented in Fig.~\ref{anomalous} allows estimating the WV: $(\sigma^\Sigma_3)^{\text{1~click}}_w=21.4\pm 4.5$.
The uncertainty corresponds to 
the width of the final pointer wavefunction in the $x$-direction.
The initial width of the pointer is $\Delta=5.8$, but for the parameters of our weak measurement procedure, the final width is $4.5$ due to a subtle narrowing effect \cite{Botero}.
We calculated this as the standard deviation $\Delta x$ of the pointer variable $x$ after the postselection, $\Delta x=\sqrt{\langle x^2 \rangle-\langle x \rangle^2}$, where
\begin{align} \label{x^2}
\langle x^2 \rangle &= \frac{\sum_{k,l=0}^{n}\binom{n}{k}\binom{n}{l}\mu^{k+l}\nu^{2n-k-l} \left((n-k-l)^2 + \Delta^2\right)\gamma_{kl}} {\sum_{k,l=0}^{n}\binom{n}{k}\binom{n}{l}\mu^{k+l}\nu^{2n-k-l}\gamma_{kl}},
\end{align}
and for our Gaussian pointer $\langle x \rangle = \left(\sigma^\Sigma_3\right)_w$.

We tested our theoretical analysis a posteriori by performing a multi-click experimental run with the same parameters $\alpha$, $\beta$ and $\Delta$, to record the $x$-distribution shown in Fig.~\ref{WV_contour_single}.
The mean value of the multi-click distribution, 18.59, is very close to the theoretical WV $(\sigma^\Sigma_3)_w=18.7$, given by Eq.~(\ref{WV7_eta}).
Since the statistical uncertainty is only 0.09, the remaining discrepancy is due to imprecision in the parameters $\alpha$, $\beta$, $\Delta$ fixing the theoretical value.
Some additional deviations are caused by imperfections in the optical elements used, e.g. the birefringent crystals (details in the Supplementary Material).
The width of the multi-click distribution in Fig.~\ref{WV_contour_single} turns out to be 4.5, in full agreement with the theoretical predictions.

To increase confidence in our results, we repeated the experiment for a few sets of different parameters leading to less anomalous WVs (and even a non-anomalous one).
All results are presented in Table \ref{tabParameters} (see the Supplementary Material for additional information).
They are also influenced by uncertainties originating from the calibration procedure.
However, when the purpose is not to find a precise numerical value of the polarization WV, but to test the single-click measurement method versus an ensemble measurement, the calibration uncertainty is irrelevant.

The experimental data shown in the table fit the theoretical predictions well, proving the possibility of measuring an anomalous WV with a single detection.

\begin{figure}
  \includegraphics[width=0.47\textwidth]{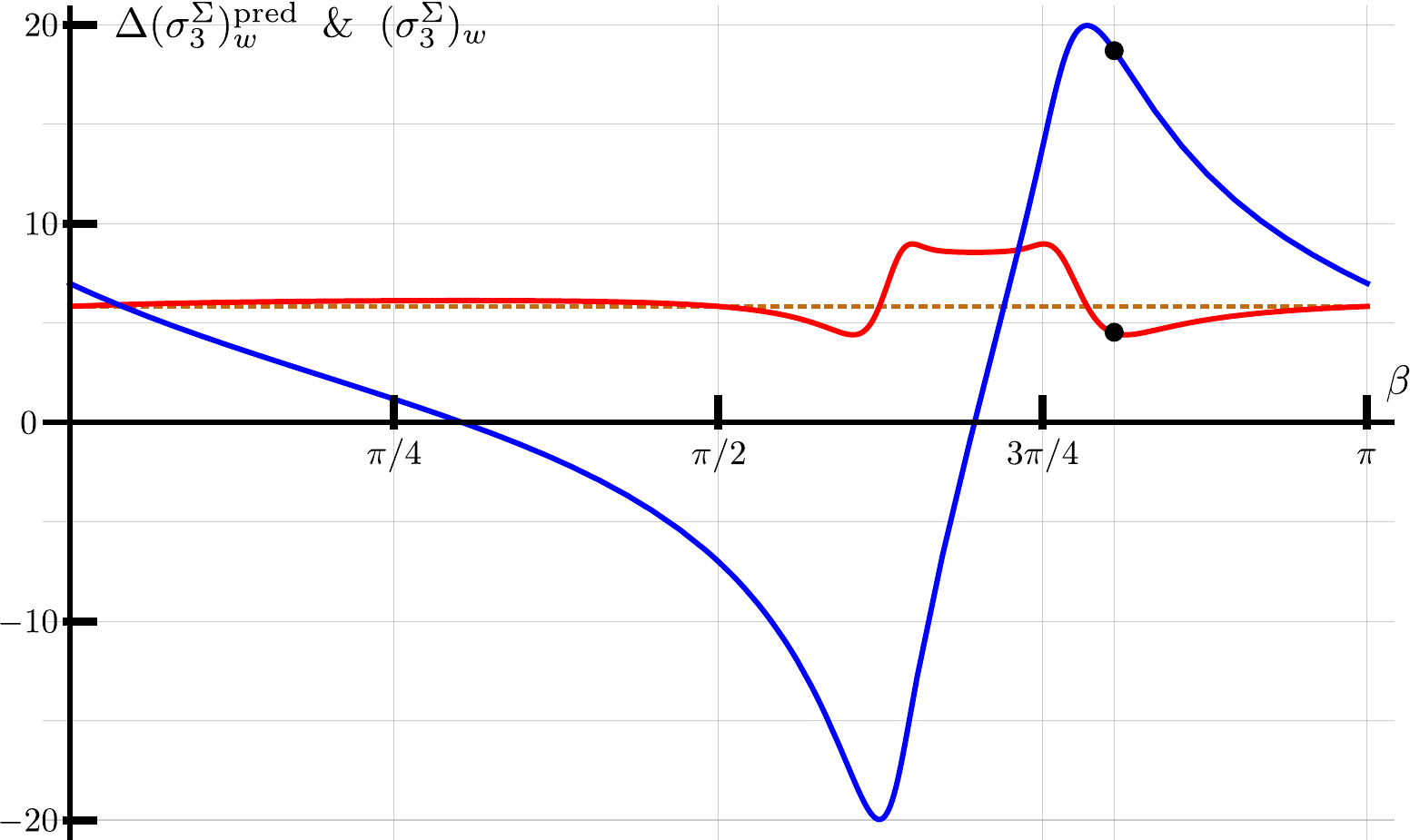}
  \caption{\textbf{Theoretical analysis.} Predicted values for the weak value $ (\sigma^\Sigma_3)_w$ (solid blue line) and the final pointer uncertainty $\Delta (\sigma^\Sigma_3)^{\text{pred}}_w$ (solid red line) with $\alpha = 0.62$, $\Delta = 5.8$ and varying $\beta$.
  The initial beam width $\Delta$ is included as a dashed brown line.
  The two black dots on the curves denote the $ (\sigma^\Sigma_3)_w = 18.7$ and $\Delta (\sigma^\Sigma_3)^{\text{pred}}_w = 4.5$ obtained for $\beta = 2.53$, i.e. the parameters of our first experimental run.}
  \label{pred_uncertainty_a_2D}
\end{figure}

Our analytical expressions (\ref{WV7_eta}) and (\ref{x^2}) allow calculating the WV and the pointer theoretical uncertainty for other pre- and postselected states.
In Fig.~\ref{pred_uncertainty_a_2D} we keep $\alpha$ and $\Delta$ as before, varying $\beta$.
In most of the cases, the final uncertainty is close to the initial beam width $\Delta$.
Our results show that, faced with a new task of $\left(\sigma^\Sigma_3\right)_w$ estimation in which somebody else fixes the pre- and postselected states of the system, a single click in our detector is capable of providing the WV with an uncertainty of the same order of magnitude as the width of the initial beam even for anomalous WVs.

In summary, our results offer a deeper understanding of the meaning of WVs, providing a significant contribution to the development of quantum measurement in the weak coupling regime \cite{AAV,Prot,DreRMP}.
On the theoretical side, our findings stress the non-statistical, single-particle nature of WVs, demonstrating how a single photon measurement can provide a WV estimate (even for anomalous WVs).
Furthermore, they prove the experimental feasibility of obtaining anomalous WVs with a single click event, suggesting a viable possibility for amplification methods without increasing the uncertainty of the measurement pointer.
This paves the way for future practical applications of the robust weak measurement paradigm.\\

\begin{table*}[]
    \centering
    \begin{tabular}{|c|c|c|c|c|c|c|c|c|c|c|}
    \hline
  \textbf{\#} & 1 & 2 & 3 & 4 & 5 & 6 & 7 & 8 & 9 & 10  \\ \hline
       & $\alpha$ & $\beta$ & $\Delta$  & $(\sigma^\Sigma_3)_w$  & $\Delta [(\sigma^\Sigma_3)_w]$ & $(\sigma^\Sigma_3)^{\text{exp}}_w$ & $\Delta_{\text{stat}}[(\sigma^\Sigma_3)^{\text{exp}}_w]$  & $(\sigma^\Sigma_3)^{\text{1~click}}_w$ & $ \Delta (\sigma^\Sigma_3)^{\text{1~click}}_w$ & $\Delta (\sigma^\Sigma_3)^{\text{pred}}_w$  \\ \hline
        (a) & 0.62 & 2.53 & 5.84 & 18.7 & 0.9 & 18.59  & 0.09 & 21.4 & 4.5 & 4.5  \\ \hline
        (b) & 0.62 & 2.53 & 3.18 & 9.8  & 0.8 & 10.51  & 0.02 & 10.9 & 3.8 & 5.0 \\ \hline
        (c) & 0.52 & 2.62 & 2.96 & 11.4 & 0.5 & 11.07  & 0.08 & 14.1 & 3.4 & 2.8 \\ \hline
        (d) & 0.52 & 0.88 & 3.09 & 1.3  & 0.4 & 0.97   & 0.05 & -2.6 & 4.6 & 3.6 \\ \hline
    \end{tabular}
 	 \caption{\textbf{Results for the various parameters of the measurement setup.} Columns 1-4 describe the preparation parameters and the corresponding weak value $(\sigma^\Sigma_3)_w$. Column 5 shows the systematic uncertainty in the experimental implementation of $(\sigma^\Sigma_3)_w$ due to the uncertainty in the preparation parameters $\alpha$, $\beta$ and $\Delta$. Column 6 shows the experimental mean values obtained by repeating the single-photon experiments. The statistical uncertainty in these experiments is shown in column 7.
 Column 8 presents the experimental weak values $(\sigma^\Sigma_3)^{\text{1~click}}_w$ extracted from a single detection event. The uncertainty in column 9 represents the quantum uncertainty of the pointer variable experimentally obtained from repeated measurements with the same parameters as the single-click experiment (see histogram in Fig.~4).
 Column 10 contains the predicted quantum uncertainty of the pointer, $\Delta(\sigma^\Sigma_3)^{\text{pred}}_w$, calculated from the parameters $\alpha$, $\beta$, and $\Delta$.} 
    \label{tabParameters}
\end{table*}{}

\section*{Methods}
Our experimental setup is shown in Fig.~\ref{setup} (further details in the Supplementary Material).
\begin{figure}
  \centering
  \includegraphics[width=0.5\textwidth]{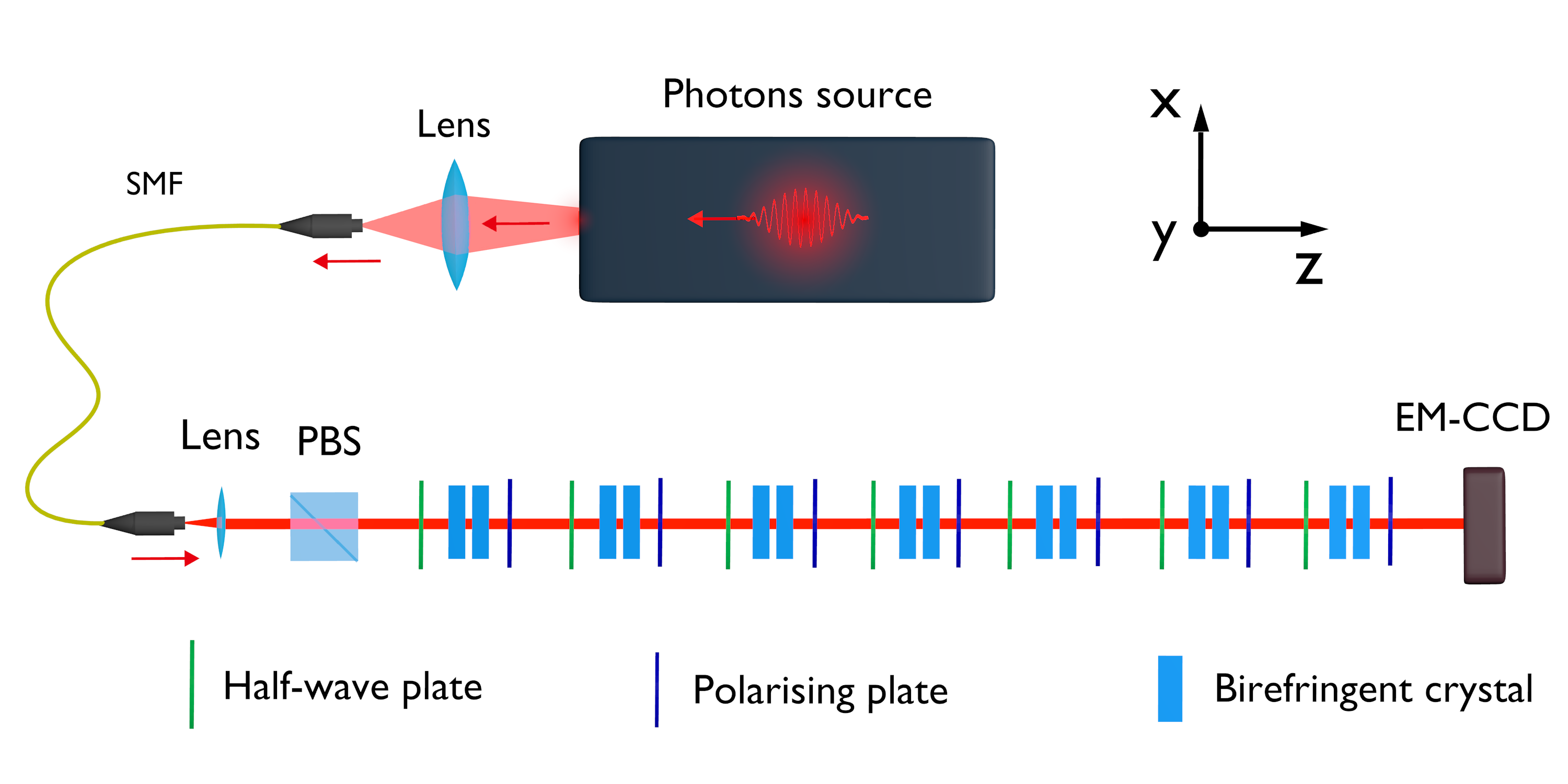}
  \caption{\textbf{Experimental setup.} Our photon source exploits type-I spontaneous parametric down conversion.
  Generated signal photons at 702 nm are spectrally-filtered, injected in a single-mode fibre and then collimated in a Gaussian mode and exploited for the experiment, while idler photons at 920 nm are detected by a Single-Photon Avalanche Detector in order to monitor the stability of the source.
  The robust weak measurement is obtained by means of the $n=7$ identical blocks put after the initial PBS. Finally, a spatial resolving detector (EM-CCD camera operating in photon counting regime) is used to determine the final position of the detected photons.
  }
  \label{setup}
\end{figure}
Photons in a multi-thermal distribution with a mean photon number per pulse  $\ll1$  are produced by type-I spontaneous parametric down-conversion (SPDC).
This guarantees a short coherence time ($\sim 150$ fs), avoiding unwanted self-interference effects due to internal reflections.
A 76 MHz mode-locked laser at 796 nm, frequency-doubled to 398 nm, is injected into a $10\times10\times5$ mm LiIO$_3$ crystal, where the SPDC takes place.
The signal photons are spectrally filtered and coupled to a single-mode fiber. At the end of the fiber, the photons are collimated in a Gaussian mode and sent to the free-space path where the robust weak measurement experiment occurs. After passing through an initial polarization beam splitter (PBS) used to suppress any residual circular polarization component, the signal photons go through $n=7$ identical blocks, each of them implementing  three steps: pre-selection, weak coupling and postselection.
Each photon enters every block in a linear polarization state, due to either the first PBS or the postselection of the previous block.
Every block begins with a quartz half-wave plate, rotating the photon polarization axis to the direction corresponding to the initial state $|\psi_\alpha\rangle$.
Then, a birefringent unit composed of a pair of birefringent crystals is responsible for the weak interaction.
The first calcite crystal, \mbox{2 mm} long, has an extraordinary ($e$) optical axis lying in the $x-z$ plane, having an angle of $\pi/4$ with respect to the $z$ direction.
This generates a spatial walk-off (of $\sim 0.2$ mm) along the $x$ direction for the horizontally-polarized photons, reducing the overlap between horizontal and vertical polarization components.
The second crystal of each unit is a $1.1$ mm long calcite crystal with the optical $e$-axis along the $y$ axis.
It generates no spatial walk-off, and its role is to compensate the temporal walk-off induced by the first crystal.
The last component of each block is a polarizing plate, postselecting the photons in the state $|\psi_\beta\rangle$.
Finally, the photons are detected by a 2D spatially resolving detector, i.e.~an Electron Multiplying CCD (EM-CCD) device able to work both in linear analog regime and in photon counting regime (details in \cite{AvellaCCD}).
To calibrate our system, we measure the position of the $|V\rangle$ polarization state, corresponding to the eigenvalue $\sigma^\Sigma_3=-7$ and then, alternatively, of the $|H\rangle$ one, corresponding to $\sigma^\Sigma_3=7$.
This allows us to define the zero point and the scale of our pointer variable: the $x$ component of the spatial wave function of the photon on our EM-CCD.


\section*{Acknowledgements}
The authors acknowledge  the European Union's Horizon 2020 and the EMPIR Participating States in the context of the projects 17FUN01 ``BeCOMe'' and 17FUN06 ``SIQUST'', the European Union's Horizon 2020 project ``Pathos'', the National Science Foundation - U.S.-Israel Binational Science Foundation Grant No. 735/18.
EC acknowledges support from the Israel Innovation Authority under project 70002 and from the Quantum Science and Technology Program of the Israeli Council of Higher Education.
JD acknowledges support from the PhD program IMPRS-QST.

\section*{Author Contributions}
IPD, MGram, MGen (responsible of the laboratories) and LV (responsible for the theoretical framework) planned the experiment, with the support of FP, AA and EC. The experimental realization was achieved (supervised by IPD, GB, MGram and MGen) by ER (leading role), FP, AA and MAdeS. The theoretical aspects were addressed by JD, EC and LV. The manuscript was prepared with inputs by all the authors. They also had a fruitful systematic discussion on the progress of the work.

\section*{Competing financial interests}
The authors declare no competing financial interests.

\end{document}